\documentstyle[prl,aps,epsf]{revtex}
\tighten
\begin{document}
\draft
\twocolumn[\hsize\textwidth\columnwidth\hsize\csname @twocolumnfalse\endcsname 
%\documentstyle[pra,aps]{revtex}
%\tightenlines

\title{Two-fluid dynamics for a Bose-Einstein condensate out of local 
equilibrium with the non-condensate}

\author{T. Nikuni}
\address{Department of Physics, Tokyo Institute of Technology, Oh-okayama,
Meguro, Tokyo 152, Japan} 

\author{E. Zaremba\cite{zaremba}}
\address{Institute for Theoretical Physics, Utrecht University, 3584 CC Utrecht,
The Netherlands}

\author{A. Griffin}
\address{Department of Physics, University of Toronto, Toronto
Ontario, Canada M5S 1A7}

\date{\today}
\maketitle

\begin{abstract}
We extend our recent work on the two-fluid
hydrodynamics of a Bose-condensed gas by including collisions 
involving both condensate and non-condensate atoms. These collisions are
essential for establishing a state of local thermodynamic equilibrium
between the condensate and non-condensate. Our theory is more general
than the usual Landau two-fluid theory, to which it reduces in the
appropriate limit, in that it allows one to 
describe situations in which a state of complete local equilibrium
between the two components has not been reached.
The exchange of atoms between the condensate and non-condensate is
associated with a new relaxational mode of the gas.
\end{abstract}
%\vskip1pc
]

Recently the authors have given a microscopic derivation of the coupled 
two-fluid hydrodynamic equations for a trapped Bose-condensed 
gas~\cite{ZGN}. In the present Letter, we report the results of a 
major extension of the ZGN work which takes into account the effects 
of collisions  between the condensate and non-condensate atoms.
These new equations of motion (which we shall call ZGN$'$) allow us 
to discuss the dynamics when the non-condensate atoms are in local 
thermal equilibrium with each
other (due to collisions between the excited atoms) but are not yet in 
complete equilibrium with the Bose condensate order parameter.
This results in the appearance of a new relaxational collective mode
related to the transfer of atoms between the condensate and 
non-condensate.  In order to bring out the new physics in a clear 
fashion, these equations are solved for a uniform gas.
A more complete derivation and discussion is given in Ref.\cite{ZGN2}.

The non-condensate atoms are described by the distribution function
$f({\bf r},{\bf p},t)$, which obeys the kinetic equation
(we take $\hbar=1$ throughout)
\begin{eqnarray}
%{\partial f({\bf r},{\bf p},t) \over \partial t} + {{\bf p} \over m} 
%\cdot \bbox{\nabla} f({\bf r},{\bf p},t) &-& \bbox{\nabla} U \cdot
%\bbox{\nabla}_{{\bf p}} f({\bf r} ,{\bf p},t) \cr
%&=& C_{12}[f] + C_{22}[f],
{\partial f \over \partial t} + {{\bf p} \over m} 
\cdot \bbox{\nabla} f - \bbox{\nabla} U \cdot
\bbox{\nabla}_{{\bf p}} f = C_{12}[f] + C_{22}[f],
\label{eq1}
\end{eqnarray}
where the effective potential
$U({\bf r},t)\equiv U_{\rm ext}({\bf r})+2g[n_c({\bf r},t)+\tilde 
n({\bf r},t)]$ involves the self-consistent Hartree-Fock (HF) mean 
field. As usual, we treat the interactions in the $s$-wave approximation
with $g=4\pi a/m$. Here $n_c({\bf r},t)$ is the condensate density and 
$\tilde n({\bf r},t)$ is the non-condensate density given by
\begin{equation}
\tilde n({\bf r},t)=\int\frac{d{\bf p}}{(2\pi)^3}f({\bf r},{\bf p},t).
\label{eq2}
\end{equation}
A kinetic equation essentially equivalent to (\ref{eq1}) in the case of
a uniform system was given in Refs.~\cite{eckern,KD}.

The two collision terms in (\ref{eq1}) are given by
\begin{eqnarray}
&&C_{22}[f] \equiv 4\pi g^2
 \int{d{\bf p}_2\over(2\pi)^3}\int{d{\bf p}_3\over(2\pi)^3}
\int d{\bf p}_4 \cr
&&\times\delta ({\bf p}+{\bf p}_2 -{\bf p}_3 -{\bf p}_4)
\delta(\tilde\varepsilon_{p}+\tilde\varepsilon_{p_2}
-\tilde\varepsilon_{p_3}-\tilde\varepsilon_{p_4}) \cr
&&\times\left[(1+f)(1+f_2)f_3f_4-ff_2(1+f_3)(1+f_4)\right]\, ,
\label{eq3}
\end{eqnarray}
\begin{eqnarray}
&&C_{12}[f]\equiv 4\pi g^2 n_c \int\frac{d{\bf p}_1}{(2\pi)^3}\int
d{\bf p}_2 \int d{\bf p}_3 \cr
&&\times  \delta(m{\bf v}_c+{\bf p}_1-{\bf p}_2-{\bf p}_3)
\delta(\varepsilon_c+\tilde\varepsilon_{p_1}
-\tilde\varepsilon_{p_2}-\tilde\varepsilon_{p_3}) \cr
&&\times  [\delta({\bf p}-{\bf p}_1)-\delta({\bf p}-{\bf p}_2)
-\delta({\bf p}-{\bf p}_3)] \cr
&&\times [(1+f_1)f_2f_3-f_1(1+f_2)(1+f_3)],
\label{eq4}
\end{eqnarray}
with $f \equiv f({\bf r, p}, t),\, f_i\equiv f({\bf r, p}_i, t)$.
Eq.~(\ref{eq4}) takes into account the fact that a
condensate atom locally has energy
$\varepsilon_c({\bf r},t)=\mu_c({\bf r},t)+\frac{1}{2}mv_c^2({\bf r},t)$
 and momentum $m{\bf v}_c$, where the condensate chemical potential 
$\mu_c$ and velocity ${\bf v}_c$ will be defined in the next
paragraph.
On the other hand, a non-condensate atom locally has the HF energy
$\tilde\varepsilon_p({\bf r},t)=\frac{p^2}{2m}+U({\bf r},t)$.
This particle-like dispersion relation means that our analysis is 
limited to finite temperatures.
It follows from this excitation spectrum that in the Landau limit (see later)
of our microscopic model, the condensate density is equal to the superfluid
density and the non-condensate density is the normal fluid density.

To complete our microscopic model, we need to have an equation of 
motion for the complex condensate order parameter 
$\Phi({\bf r},t)\equiv\sqrt{n_c({\bf r},t)}e^{i \theta({\bf r},t)}$.
This equation can be rewritten in terms of $n_c({\bf r},t)$ and the 
condensate velocity ${\bf v}_c=\bbox{\nabla}\theta({\bf r},t)/m$ as
\begin{mathletters}
\begin{eqnarray}
{\partial n_c \over \partial t} + \bbox{\nabla}\cdot(n_c{\bf v}_c)&=& 
-\Gamma_{12}[f]\,, \label{eq5a} \\
m\left({\partial\over\partial t}+{\bf v}_c\cdot 
\bbox{\nabla}\right) {\bf v}_c&=&-\bbox{\nabla}\mu_c,
\label{eq5b}
\end{eqnarray}
\label{eq5}
\end{mathletters}

\noindent
where the condensate chemical potential (in the Thomas-Fermi 
approximation) is given by
\begin{equation}
\mu_c({\bf r}, t) =U_{\rm ext}({\bf r})+
g[n_c({\bf r}, t)+2\tilde{n}({\bf r}, t)]\, .
\label{eq6}
\end{equation}
The ``source" term $\Gamma_{12}[f]$ in (\ref{eq5a}) is defined in 
terms of the $C_{12}$ collision term in (\ref{eq4}) as
\begin{equation}
\Gamma_{12}[f]\equiv\int\frac{d{\bf p}}{(2\pi)^3}C_{12}[f({\bf r},
{\bf p},t)].
\label{eq7}
\end{equation}
We observe that $C_{12}$ collisions do not conserve the number of atoms
in the condensate.

The detailed derivation~\cite{ZGN2} 
of Eqs.~(\ref{eq1}-\ref{eq7}) is based on a 
field-theoretic formulation of an interacting Bose fluid 
with a Bose broken symmetry. 
Closely related work can be found in Refs.~\cite{zoller}
and~\cite{stoof}.
The quantum field operators 
$\hat\psi({\bf r})$ are split into a condensate
($\Phi\equiv\langle\hat\psi\rangle$) and non-condensate ($\tilde\psi$) 
part. The key approximation made in obtaining (\ref{eq1}) and 
(\ref{eq5}) is 
the neglect of the anomalous pair correlations such as
$\langle \tilde\psi({\bf r})\tilde\psi({\bf r})\rangle$,
which we shall refer to as the Popov approximation.
In~\cite{ZGN}, the $C_{12}$ collisions (involving
one condensate atom) were not included and thus
the source term $\Gamma_{12}$ in (\ref{eq5a}) was not present.
In the extended set of ZGN$'$ hydrodynamic equations 
which follow from (\ref{eq1}) and (\ref{eq5}),
$\Gamma_{12}$ plays a crucial role.  Above the Bose-Einstein transition
temperature ($T_{\rm BEC}$), where the Bose condensate order 
parameter vanishes, the kinetic equation (\ref{eq1}) only
involves the Uehling-Uhlenbeck collision term $C_{22}$ given by 
(\ref{eq3}). This kinetic equation has been discussed extensively 
in recent years~\cite{smerzi}.

In this Letter, we restrict ourselves to the region in which $C_{22}$ 
collisions are sufficiently rapid to justify the assumption that the 
excited-atom distribution function is approximately described by the 
local equilibrium Bose distribution 
\begin{equation}
\tilde f({\bf r},{\bf p},t)=
\frac{1}{e^{\beta[\frac{1}{2m}({\bf p}-m{\bf v}_n)^2+U-\tilde \mu]}-1}
\, .
\label{eq8}
\end{equation}
Here, the temperature parameter $\beta$, normal fluid velocity 
${\bf v}_n$, chemical potential $\tilde \mu$, and mean field $U$ are all
functions of ${\bf r}$ and $t$. It is important to appreciate that the
local non-condensate chemical potential $\tilde \mu$
which appears in (\ref{eq8}) is distinct from the local
condensate chemical potential $\mu_c$, as defined in (\ref{eq6}).
One may immediately verify that $\tilde f$ satisfies 
$C_{22}[\tilde f]=0$, independent of the value of $\tilde \mu$.
In contrast, one finds from (\ref{eq4}) that, in general, 
$C_{12}[\tilde f]\neq 0$. This means that even if the excited atoms 
are in dynamic local equilibrium described by (\ref{eq8}), the 
source term $\Gamma_{12}[f]$ in (\ref{eq5}) will, in general, be
finite.  More specifically, we have (see also Ref.~\cite{zoller})
\begin{equation}
\Gamma_{12}[\tilde f]=- \left\{1-e^{-\beta[\tilde\mu-
\frac{1}{2}m({\bf v}_n-{\bf v}_c)^2-\mu_c]}\right\}
\frac{n_c}{\tau_{12}},
\label{eq9}
\end{equation}
where we have introduced a collision time associated with the 
$C_{12}$ term in (\ref{eq4}),
\begin{eqnarray}
&&\frac{1}{\tau_{12}}\equiv4\pi g^2
\int\frac{d{\bf p}_1}{(2\pi)^3}\int \frac{d{\bf p}_2}{(2\pi)^3}
\int d{\bf p}_3 (1+\tilde f_1)\tilde f_2 \tilde f_3 \cr
&& \ \ \times\delta(m{\bf v}_c+{\bf p}_1-{\bf p}_2-{\bf p}_3)
\delta(\varepsilon_{c}+\tilde\varepsilon_{p_1}
-\tilde\varepsilon_{p_2}-\tilde\varepsilon_{p_3})\, .
\label{eq10}
\end{eqnarray}
We note that $\Gamma_{12}[\tilde f]$ in (\ref{eq9}) vanishes when
$\tilde\mu=\mu_c+\frac{1}{2}m({\bf v}_n-{\bf v}_c)^2$.
However, as we shall see, a state of complete local equilibrium cannot
be treated simply by setting $\Gamma_{12}=0$, which was the implicit
assumption made in deriving the Landau two-fluid equations in 
earlier work~\cite{ZGN,KD}.

With the assumption $f \simeq \tilde f$,
one can derive hydrodynamic equations for the 
non-condensate by taking moments of (\ref{eq1}) in the usual way.
These are the analogue of the condensate equations in (\ref{eq5}).
Linearizing around the static thermal equilibrium solutions,
these hydrodynamic equations are given by
\begin{mathletters}
\begin{eqnarray}
{\partial \delta\tilde{n}\over\partial t}
&=&-\bbox{\nabla}\cdot(\tilde{n}_0\delta{\bf v}_n)
+\delta\Gamma_{12}\,, \label{eq11a}\\
m\tilde{n}_0\frac{\partial\delta{\bf v}_n}{\partial t}
&=& -\bbox{\nabla}\delta\tilde{P}-\delta\tilde n\bbox{\nabla}U_0 \cr
&& \ \ \ \ \ \ \ \ \ \ -2g\tilde n_0 \bbox{\nabla}
(\delta\tilde n+\delta n_c)\,,\label{eq11b} \\ 
\frac{\partial \delta\tilde P}{\partial t} &=&-\frac{5}{3}
\tilde P_0\bbox{\nabla}\cdot\delta{\bf v}_n-\delta{\bf v}_n\cdot
\bbox{\nabla}\tilde P_0 \cr
&& \ \ \ \ \ \ \ \ \ \ \ \ \ \ \ \ \ \ \ 
-\frac{2}{3}gn_{c0}\delta\Gamma_{12}
\label{eq11c}\,.
\end{eqnarray}
\label{eq11}
\end{mathletters}
One also has
\begin{equation}
\tilde n({\bf r},t)=\int\frac{d{\bf p}}{(2\pi)^3}\tilde 
f({\bf r},{\bf p},t)= \frac{1}{\Lambda^3}g_{3/2}(z)\, , 
\label{eq12}
\end{equation}
\begin{eqnarray}
\tilde P({\bf r},t)&=&\int\frac{d{\bf p}}{(2\pi)^3} {p^2 \over 3m}
\tilde f({\bf r},{\bf p},t)\big \vert_{{\bf v}_n=0} \nonumber \\
&=& {1\over \beta\Lambda^3}g_{5/2}(z)\,,
\label{eq13}
\end{eqnarray}
where $z\equiv e^{\beta(\tilde\mu-U)}$ is the local fugacity,
$\Lambda\equiv\sqrt{2\pi/mk_{\rm B}T}$ is the local thermal de Broglie
wavelength and $g_n(z)\equiv\sum_{l=1}^{\infty}z^l/l^n$.
In static equilibrium (denoted by $0$), we of course have
${\bf v}_{n0}={\bf v}_{c0}=0$ and $\mu_{c0}=\tilde\mu_0$.
Thus it follows that $\Gamma_{12}[\tilde f^0]=0$.
The analogous linearized condensate equations of motion are
\begin{mathletters}
\begin{eqnarray}
{\partial \delta n_c \over \partial t} &=& -
\bbox{\nabla}\cdot(n_{c0}\delta{\bf v}_c)
-\delta\Gamma_{12}, \label{eq14a} \\
m\frac{\partial \delta {\bf v}_c}{\partial t}
&=&-g\bbox{\nabla}(\delta n_c+2\delta \tilde n).
\label{eq14b}
\end{eqnarray}
\label{eq14}
\end{mathletters}

\noindent
Finally, the source term $\delta\Gamma_{12}$ in these equations can be 
expressed in terms of the fluctuation in the chemical
potential difference $\mu_{\rm diff}\equiv\tilde\mu-\mu_c$,
\begin{equation}
\delta\Gamma_{12}=-\frac{\beta_0n_{c0}}{\tau^0_{12}}
\delta\mu_{\rm diff},
\label{eq15}
\end{equation}
where $\tau_{12}^0$ is the equilibrium collision time obtained from
(\ref{eq10}) with ${\bf v}_c=0$, $\varepsilon_c=\mu_{c0}$ and
$\tilde f$ equal to the absolute equilibrium Bose distribution.
%\begin{equation}
%\tilde f_i^0\equiv\frac{1}{e^{\beta_0(\frac{p_i^2}{2m}+U_0
%-\mu_{c0})}-1}.
%\label{eq16}
%\end{equation}
We note that adding (\ref{eq11a}) and (\ref{eq14a}) gives the 
usual continuity equation for the total density.

We now turn to a discussion of our linearized hydrodynamic equations
given by (\ref{eq11})-(\ref{eq15}) for a {\it uniform} Bose-condensed 
gas ($U_{\rm ext}({\bf r})=0$). This means $\tilde n_0$, 
$n_{c0}$ and $\tilde P_0$ are independent of position.
By straightforward calculations~\cite{ZGN2}, one can then reduce our 
two-fluid equations to three coupled equations of motion for the 
three variables $\delta{\bf v}_c$, $\delta{\bf v}_n$ and 
$\delta\mu_{\rm diff}$:
\begin{mathletters}
\begin{eqnarray}
&&m\frac{\partial^2\delta{\bf v}_c}{\partial t^2}
=gn_{c0}\bbox{\nabla}(\bbox{\nabla}\cdot\delta{\bf v}_c)
+2g\tilde n_0 \bbox{\nabla}(\bbox{\nabla}\cdot\delta{\bf v}_n) \cr
&& \ \ \ \ \ \ \ \ \ \ \ \ \ \ \ \ \ \ \ \ \ \  
+\frac{\beta_0gn_{c0}}{\tau^0_{12}}{\bbox{\nabla}\delta\mu_{\rm diff}}
\,, \label{eq17a} \\
&&m\frac{\partial^2\delta{\bf v}_n}{\partial t^2}
=\left(\frac{5\tilde P_0}{3\tilde n_0}+2g\tilde n_0\right)
\bbox{\nabla}(\bbox{\nabla}\cdot\delta{\bf v}_n) \cr
&& \ \ \ \ \ \ \ +2gn_{c0}\bbox{\nabla}(\bbox{\nabla}\cdot\delta{\bf v}_c)
-\frac{2n_{c0}}{3\tilde n_0}\frac{\beta_0gn_{c0}}{\tau^0_{12}}\bbox{\nabla}
\delta\mu_{\rm diff}\,, \label{eq17b} \\
&&\frac{\partial\delta\mu_{\rm diff}}{\partial t}=gn_{c0}
\left(\frac{2}{3}\bbox{\nabla}\cdot\delta{\bf v}_n-\bbox{\nabla}\cdot
\delta{\bf v}_s\right)
-\frac{\delta\mu_{\rm diff}}{\tau_{\mu}}\,.\label{eq17c}
\end{eqnarray}
\label{eq17}
\end{mathletters}

\noindent
Here the relaxation time for the chemical potential difference 
($\mu_{\rm diff}$) due to $C_{12}$
collisions between the condensate and non-condensate atoms is given by
the expression~\cite{ZGN2}
\begin{equation}
{1\over \tau_\mu} \equiv {\beta_0 g n_{c0} \over \tau^0_{12}}
\left ( {{5\over 2} \tilde P_0 + 2g\tilde n_0 
n_{c0} + {2\over 3} \tilde\gamma_0 g n_{c0}^2 \over {5\over 2} 
\tilde\gamma_0 \tilde P_0 - {3\over 2} g\tilde n_0^2} -1 \right ),
\label{eq18}
\end{equation}
where we have introduced the dimensionless function
$\tilde\gamma_0\equiv (g\beta_0/\Lambda_0^3)g_{1/2}(z_0)$.
If we simply omit the terms involving 
$\delta\mu_{\rm diff}$ in (\ref{eq17a}) and (\ref{eq17b}),
we are left with the two coupled ZGN equations for $\delta{\bf v}_n$ 
and $\delta{\bf v}_c$ given in Ref.\cite{GZ}. We see that  our new 
generalized ZGN$'$ equations give rise to a coupling between
$\delta{\bf v}_n$ and $\delta{\bf v}_c$ and the local variable
$\delta\mu_{\rm diff}$, which describes the relative fluctuation in 
the chemical potentials of the two components.
This is the most important new result in the present Letter.

It is convenient to solve the three coupled equations in (\ref{eq17})
by introducing velocity potentials $\delta
{\bf v}_c \equiv \bbox{\nabla} \phi_c$ and
$\delta {\bf v}_n \equiv \bbox{\nabla} \phi_n$,
and looking for plane wave solutions. We obtain from (\ref{eq17c})
\begin{equation}
(1-i\omega\tau_{\mu})\delta \mu_{\rm diff} = 
gn_{c0}\tau_\mu \left (
\phi_c - {2\over 3}\phi_n \right ) k^2.
\label{eq19}
\end{equation}
Inserting this expression into (\ref{eq17a}) and (\ref{eq17b}) 
gives two coupled equations for $\phi_n$ and $\phi_{c}$,
\begin{mathletters}
\begin{eqnarray}
&&m\omega^2 \phi_c = gn_{c0}\left [ 1 - {\beta_0 gn_{c0} \tau_\mu
\over \tau^0_{12} (1-i\omega\tau_\mu)} \right ] k^2 \phi_c \cr
&&\qquad+2g\tilde n_0 \left [ 1 + {\beta_0 gn_{c0}\tau_\mu \over 3
\tau^0_{12}(1-i\omega\tau_\mu) } {n_{c0} \over \tilde n_0} 
\right ] k^2 \phi_n \,, \label{eq20a} \\
&&m\omega^2 \phi_n = 
2gn_{c0} \left[ 1 + {\beta_0 gn_{c0}\tau_\mu \over 3
\tau^0_{12}(1-i\omega\tau_\mu) } {n_{c0} \over \tilde n_0} 
\right] k^2 \phi_c \cr
&&\qquad+\left[ {5\tilde P_0 \over 3\tilde n_0} + 2g\tilde
n_0 - {4\beta_0 (gn_{c0})^2 \tau_\mu \over 9 \tau^0_{12}
(1-i\omega\tau_\mu) } {n_{c0}\over \tilde n_0} \right] k^2 \phi_n 
\,. \label{eq20b}
\end{eqnarray}
\label{eq20}
\end{mathletters}

\noindent
Taking the limit $\omega\tau_{\mu}\gg 1$,
one finds that $\delta\mu_{\rm diff}$ is decoupled from the velocity 
potentials $\phi_{c,n}$ and we recover the ZGN results~\cite{GZ}.
As expected, in the extreme limit $\omega\tau_{\mu}\gg1$, the effect of
$C_{12}$ collisions are negligible and one can simply omit 
$\Gamma_{12}$.

In the opposite limit $\omega \tau_\mu \to 0$, the equations in 
(\ref{eq20}) yield
%\begin{mathletters}
%\begin{eqnarray}
%m\omega^2 \phi_c &=& gn_{c0}\Bigg ( 1 - {\beta_0 gn_{c0} \tau_\mu
%\over \tau^0_{12} } \Bigg ) k^2 \phi_c
%+ 2g\tilde n_0 \left ( 1 + {\beta_0 gn_{c0}\tau_\mu \over 3 \tau^0_{12} } 
%{n_{c0} \over \tilde n_0} \right ) k^2 \phi_n\label{eq92a} \\ 
%m\omega^2 \phi_n &=& \left( {5\tilde P_0 \over 3\tilde n_0} + 2g\tilde
%n_0 - {4\beta_0 (gn_{c0})^2 \tau_\mu \over 9 \tau^0_{12} } 
%{n_{c0}\over \tilde n_0} \right) k^2 \phi_n 
%+ 2gn_{c0} \left ( 1 + {\beta_0 gn_{c0}\tau_\mu \over 3
%\tau^0_{12}} {n_{c0} \over \tilde n_0} \right ) k^2 \phi_c\,. 
%\label{eq92b}
%\end{eqnarray}
%\label{eq92}
%\end{mathletters}
%
%\noindent
%This coupled set of equations has
two phonon-like solutions, $\omega_{1,2}=u_{1,2}k$, where the
velocities are given by the roots of the equation $u^4-Au^2+B=0$. 
It can be shown that the coefficients $A$ and $B$
are in {\it exact} agreement with the analogous coefficients obtained
from the usual Landau two-fluid equations~\cite{GZ,ho}.  The latter 
theory uses quite different thermodynamic variables from those
used in the present formulation,
and the explicit proof of this equivalence requires
a lengthy (but straightforward) calculation.
It also turns out that the first and second sound velocities ($u_1$ and $u_2$)
given by these results (valid for $\omega\tau_{\mu}\ll 1$) are numerically
very close to the velocities given by the ZGN approximation
(valid for $\omega\tau_{\mu}\gg 1$).
The small differences involve terms of order $g^2$
and thus were not picked up in the comparison given in Ref.~\cite{GZ}.

The interesting feature of the linearized ZGN$'$ equations in 
(\ref{eq19}) and (\ref{eq20}) is the existence of a new mode, 
associated with the condensate and non-condensate
being out of equilibrium ($\delta\mu_c\neq\delta\tilde\mu$).
To a good approximation (and exact in the $k\to 0$ limit),
it corresponds to a mode in which
$\delta{\bf v}_n=\delta{\bf v}_c=0$, with a frequency given by
$\omega=-i/\tau_{\mu}$. In the ZGN limit ($C_{12}=0$), this reduces 
to a zero frequency mode. In the Landau limit ($C_{12}$ large), 
it is a heavily damped relaxational mode.
According to (\ref{eq20}), this equilibration process also gives rise 
to a damping of second sound whose magnitude
relative to the mode frequency is peaked at 
$\omega\tau_{\mu}\simeq 1$.

In Fig.~1, we plot the reciprocals of various 
relaxation times involved in our linearized
ZGN$'$ equations, as a function of temperature.
We note that $\tau_{12}^0$ goes to zero at $T_{\rm BEC}$ and is much smaller than
the mean collision time expected for a Maxwell-Boltzmann gas.
The extra factors multiplying $1/\tau^0_{12}$ in 
(\ref{eq18}) ensure that $1/\tau_{\mu}$
starts to decrease as we approach $T_{\rm BEC}$ from below.
This is the expected ``critical slowing down" seen in all second
order phase transitions involving an order parameter.
Our HF mean-field approximation is inadequate in the critical region
close to $T_{\rm BEC}$, and leads to a spurious finite limiting value of
$n_{c0}$ at $T_{\rm BEC}$~\cite{GZ}. This removes the divergence in
$1/\tau^0_{12}$, which 
is also why $\tau_{\mu}$ in Fig.~1 is finite at $T_{\rm BEC}$.
If we simply put $n_{c0}=0$ in our calculations, we would 
find that both $\tau_{\mu}$ and $1/\tau_{12}^0$ would diverge at 
$T_{\rm BEC}$.

Our discussion in the present Letter is based on the assumption that 
$C_{22}$ collisions produce the local equilibrium distribution 
$\tilde f$ in (\ref{eq8}).
As a result, our hydrodynamic equations (\ref{eq11}) do not explicitly depend on
a relaxation time associated with $C_{22}$. This could be included 
using the standard Chapman-Enskog approach to deal with the deviation 
of $f$ from the local equilibrium function $\tilde f$~\cite{KD,NG}.
However, an estimate of the collision time associated with $C_{22}$ 
is given by the scattering out term in (\ref{eq3}),
\begin{eqnarray}
&&\frac{1}{\tau^0_{22}}=\frac{4\pi g^2}{\tilde n_0}
\int\frac{d{\bf p}_1}{(2\pi)^3}\int\frac{d{\bf p}_2}{(2\pi)^3}
\int\frac{d{\bf p}_3}{(2\pi)^3}\int d{\bf p}_4  \cr
&& \ \ \ \ \ \times \delta({\bf p}_1+{\bf p}_2-{\bf p}_3-{\bf p}_4)
\delta(\tilde\varepsilon_{p_1}+\tilde\varepsilon_{p_2}-
\tilde\varepsilon_{p_3} -\tilde\varepsilon_{p_4}) \cr
&& \ \ \ \ \ \times \tilde f_1^0 \tilde f_2^0(1+ \tilde f_3^0)
(1+ \tilde f_4^0).
\label{eq21}
\end{eqnarray}
This  collision time is plotted in Fig.~1,
both above and below $T_{\rm BEC}$.  The fact that 
$\tau^0_{22}\ll\tau_{\mu}$ at temperatures $T\gtrsim0.8~T_{\rm BEC}$
is very important, since it allows for the possibility that
the non-condensate atoms are in local equilibrium with each other
but not with the condensate.
Above the transition, the value for $\tau_{22}^0$
we obtain is in close agreement with the collision
time obtained in Ref.~\cite{smerzi}.

These results for the relaxation times are quite interesting in their own right.
The divergence at $T_{\rm BEC}$ (see above remarks)
is a consequence of the Bose distribution function
being used for $\tilde f_i^0$ in (\ref{eq10}) and (\ref{eq21}).
If a Maxwell-Boltzmann distribution function were used, the ``divergence''
shown in Fig.~1 is removed
(the importance of calculating collision times using the correct Bose distribution
has also been noted in Ref.~\cite{zoller}).
Moreover, we see that one should not use the classical gas approximation
for the collision times when determining the 
cross-over from a collisionless (or mean-field) to hydrodynamic regimes.
The results in Fig.~1 imply that the hydrodynamic domain is
much easier to reach
at finite temperatures than might have been expected, since the collision time
can be much {\it smaller}
than the analogous classical gas collision time.
%A related point was made in Ref.~\cite{smerzi} above $T_{\rm BEC}$.

In summary, starting from a microscopic model, we find that the 
dynamics of a
Bose-condensed gas at finite temperatures can be divided into {\it three}
distinct regimes:
(1) The collisionless regime in which no collision terms are included 
in the kinetic equation ($C_{12}=C_{22}=0$).
(2) An intermediate regime in which $C_{22}$ collisions 
between the excited atoms
establish local thermal equilibrium ($\omega\tau_{22}^0 \ll 1$)
but the $C_{12}$ collisions do not keep
the condensate in equilibrium with the non-condensate~\cite{eckern,MY}.
The relaxation time $\tau_{\mu}$ for this equilibration is
found to be much larger than the collision time $\tau^0_{22}$ for reaching local
equilibrium in the non-condensate (see Fig.~1).
(3) Complete local equilibrium of the condensate and non-condensate, which arises
in the limit $\omega\tau_{\mu}\ll1$.
This is the regime conventionally described by the Landau two-fluid 
equations \cite{GZ,ho}. As stated earlier, the ZGN$'$ 
equations exactly 
reproduce the results of the Landau equations, even though
the local dynamical variables used are quite different in the two 
approaches.

In this Letter, 
we have only discussed the normal modes of our linearized equations
in a uniform Bose gas, but similar conclusions
are obtained for trapped gases~\cite{ZGN2}.
Our general equations can also be used to discuss
the growth and decay of atomic condensates \cite{zoller,DMS},
taking the dynamics of the non-condensate into account.

We thank H.T.C. Stoof for first emphasizing the existence of an 
additional mode in the ZGN equations. T.N. is supported by a JSPS 
fellowship, while E.Z. and A.G. are supported by research grants from 
NSERC.
E.Z. also acknowledges the financial support of the Dutch Foundation FOM.

\noindent

\begin{figure*}
\begin{caption}
{
Various collision and relaxation times for a uniform Bose gas
as a function of temperature, for $gn=0.1k_{\rm B}T_{\rm BEC}$. 
The values are normalized to the classical collision time
$\tau_{\rm cl}^{-1}=n\sigma(16 k_B T/\pi m)^{1/2}$
at $T=T_{\rm BEC}$, obtained
from (\ref{eq21}) using a Maxwell-Boltzmann distribution for $f$.
Here, $\sigma = 8\pi a^2$ is the atomic cross-section.
The temperature scale changes abruptly above $T_{\rm BEC}$.
The present calculations are not valid in the
very low temperature region.}
\end{caption}
\label{fig1}
\end{figure*}
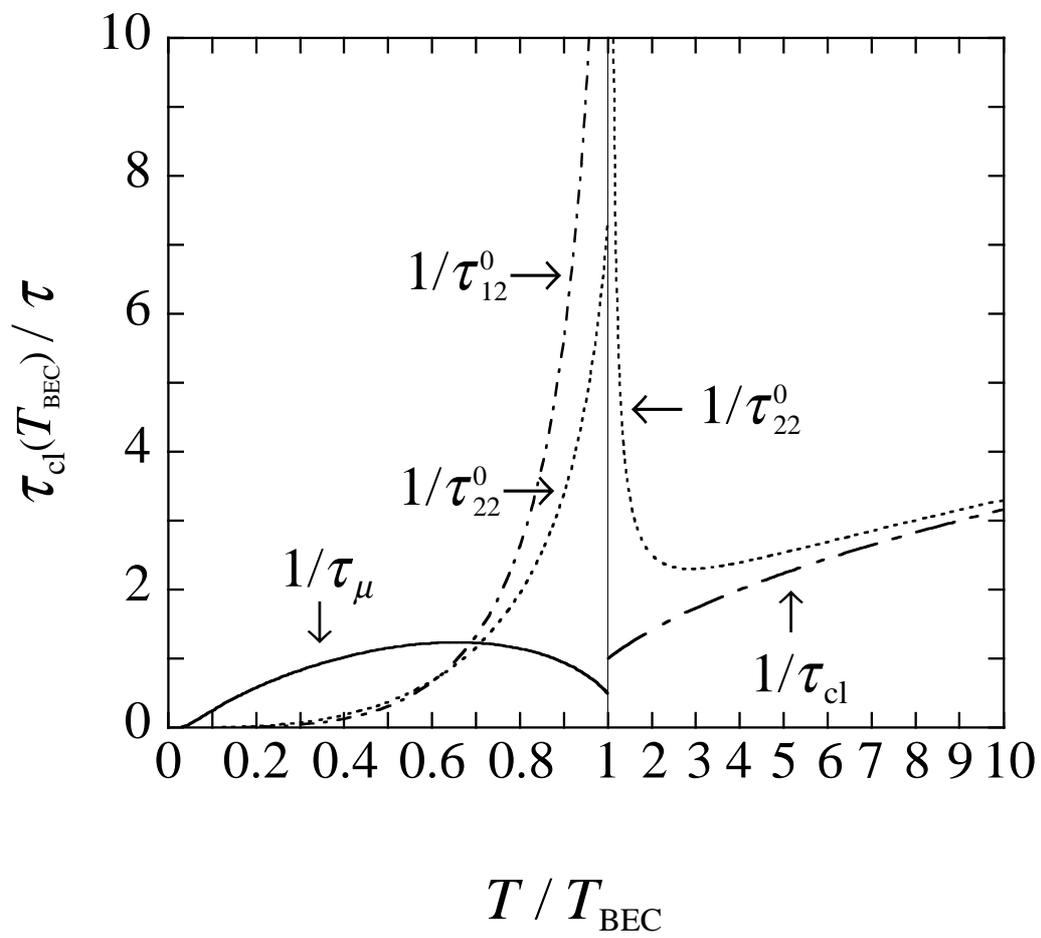
\end{document}